# Jet power, intrinsic $\gamma$-ray luminosity, and accretion in jetted AGNs

Yongyun Chen(陈永云),[1]★ Qiusheng Gu(顾秋生),[2]★ Junhui Fan(樊军辉),[3] Xiaoling Yu(俞效龄),[1] Nan Ding(丁楠),[4] Xiaotong Guo(郭晓通),[5] and Dingrong Xiong(熊定荣)[6]

[1]*College of Physics and Electronic Engineering, Qujing Normal University, Qujing 655011, P. R. China*
[2]*School of Astronomy and Space Science, Nanjing University, Nanjing 210093, P. R. China*
[3]*Center for Astrophysics, Guang zhou University, Guang zhou 510006, P. R. China*
[4]*School of Physical Science and Technology, Kunming University, 650214, Kun ming, P. R. China*
[5]*School of Mathematics and Physics, Anqing Normal University, 246011, An qing, P. R. China*
[6]*Yunnan Observatories, Chinese Academy of Sciences, Kunming 650011, P. R. China*

## ABSTRACT

The correlation between the kinetic jet power $P_{\rm jet}$, intrinsic $\gamma$-ray luminosity ($L^{\rm int}$), and accretion ($L_{\rm disc}$) may reveal the underlying jet physics in various black hole systems. We study the relation between kinetic jet power, intrinsic $\gamma$-ray luminosity, and accretion by using a large sample of jetted active galactic nuclei (AGNs), including flat-spectrum radio quasars (FSRQs), BL Lacertae objects (BL Lacs), $\gamma$-ray narrow-line Seyfert 1 galaxies ($\gamma$NLS1s), and radio galaxies. Our main results are as follows: (1) The slope indices of the relation between $P_{\rm jet}$ and $L^{\rm int}$ are $0.85 \pm 0.01$ for the whole sample, $0.70 \pm 0.02$ for the FSRQs, $0.83 \pm 0.03$ for the BL Lacs, $0.68 \pm 0.11$ for the $\gamma$NLS1s, and $0.93 \pm 0.09$ for the radio galaxies, respectively. The jets in $\gamma$NLS1s and radio galaxies almost follow the same $P_{\rm jet}$–$L^{\rm int}$ correlation that was obtained for Fermi blazars. (2) The slope indices of the relation between $L^{\rm int}$ and $L_{\rm disc}$ are $1.05 \pm 0.02$ for the whole sample, $0.94 \pm 0.05$ for the FSRQs, $1.14 \pm 0.05$ for the BL Lacs, and $0.92 \pm 0.18$ for the $\gamma$NLS1s, respectively. The $\gamma$NLS1s and radio galaxies almost also follow the $L^{\rm int}$–$L_{\rm disc}$ correlation derived for Fermi blazars. (3) The jet power is larger than the luminosity of accretion discs for almost all jetted AGNs. Jet power depends on both the Eddington ratio and black hole mass. We obtain $\log P_{\rm jet} \sim (1.00 \pm 0.02)\log L_{\rm disc}$ for the whole sample, which is consistent with the theoretically predicted coefficient. These results may imply that the jets of jetted AGNs are powered by the Blandford–Znajek mechanism.

## 1 INTRODUCTION

Blazars are a special subclass of active galactic nuclei (AGNs) whose relativistic jets point towards the observer. According to the equivalent width (EW) of the broad emission lines, blazars are divided into two subclasses: flat-spectrum radio quasars (FSRQs) and BL Lacertae objects (BL Lacs). EWs greater than 5 Å are FSRQs, otherwise it is BL Lacs (Urry & Padovani 1995). Subsequently, some authors introduced some more physical classifications. Ghisellini et al. (2011) found that FSRQs and BL Lacs can be separated by the ratio of the luminosity of the broad-line region (BLR) to the Eddington luminosity, and the FSRQs have $L_{\rm BLR}/L_{\rm Edd} \geq 5 \times 10^{-4}$, and BL Lacs is less than this value. Sbarrato, Padovani & Ghisellini (2014) found that FSRQs have $L_{\rm BLR}/L_{\rm Edd} \geq 10^{-3}$, and BL Lacs have $L_{\rm BLR}/L_{\rm Edd} < 10^{-3}$. These authors suggest that they may reflect changes in the accretion model.

Radio galaxies are considered to be misaligned blazars: blazars have a small viewing angle, while radio galaxies have a large viewing angle (Meyer et al. 2011). According to radio morphology, radio galaxies are usually divided into two subclasses: FR I and FR II (Fanaroff & Riley 1974). According to the unified model, FSRQs and FR II radio galaxies are unified, while BL Lacs and FR I radio galaxies are unified (Urry & Padovani 1995). Chen et al. (2015b) found that FSRQs and FRII radio galaxies are in the radiation-pressure-dominated regime, while BL Lacs and FR I radio galaxies are in the gas-pressure-dominated regime. Sbarrato et al. (2014) found a tight connection between the $\gamma$-ray luminosity and the luminosity of the BLR using small samples of Fermi blazars and radio galaxies. These results may imply that the blazars and radio galaxies have similar jets and accretion properties.

The radio-loud narrow-line Seyfert 1 galaxies (RLNLS1s) is another important subclass of the unified model of AGNs (Foschini 2017). Some authors have found a close relationship between Fermi blazars and RLNLS1s. The jet power of FSRQs and RLNLS1s depends on the black hole mass, which implies that the accretion discs of FSRQs and RLNLS1s are dominated by the radiation pressure (Foschini 2011; Chen & Gu 2019). The physical properties of RLNLS1s are similar to that of Fermi blazars (e.g. Foschini et al. 2015; Sun et al. 2015; Berton et al. 2018; Paliya et al. 2019). Chen et al. (2021a) found that there is a weak anticorrelation between synchrotron peak frequency and peak luminosity for both Fermi blazars and RLNLS1s, which suggests that the RLNLS1s belong to the Fermi blazar sequence.

The formation mechanism of relativistic jets has always been a hot issue in astrophysical research. At present, there are three main mech-anisms for the formation of jets. The first is the Blandford–Znajek (BZ) mechanism: the jets extract the rotational energy of the black hole and accretion disc (Blandford & Znajek 1977). The BZ jet power depends on the spin of the black hole, and the square of the magnetic flux threading the black hole horizon. The second is the Blandford–Payne (BP) mechanism: jet extracts only the rotational energy of the accretion disc (Blandford & Payne 1982), and a black hole is not necessary. In both cases, it should be sustained by matter accreting on to the black hole, leading one to expect a relation between accretion and jet power (Maraschi & Tavecchio 2003). Several authors have demonstrated this correlation using a small sample (e.g. Rawlings & Saunders 1991; Cao & Jiang 1999; Ghisellini, Tavecchio & Ghirlanda 2009; Gu, Cao & Jiang 2009; Ghisellini et al. 2010, 2011; Sbarrato et al. 2012, 2014; Chen et al. 2015a, b). The third is Hybrid models: a mixture of BZ and BP mechanisms (Meier 2001; Garofalo, Evans & Sambruna 2010). Garofalo et al. (2010) used the hybrid model to speculate on the observed differences in AGNs with relativistic jets.

★ E-mail: ynkmcyy@yeah.net (YC); qsgu@nju.edu.cn (QG)

Relativistic jets are ubiquitous in the universe and have been observed in various black hole systems ranging from stellar mass to supermassive black holes. One outstanding question is how the jet physics scale with mass from stellar to supermassive black hole. There is evidence to suggest that jets behave in similar ways in blazars, low-luminosity active galactic nuclei (LLAGNs), black hole X-ray binaries (XRBs), and $\gamma$-ray bursts (GRBs) (e.g. Merloni, Heinz & di Matteo 2003; Falcke, Körding & Markoff 2004; Nemmen et al. 2012; Zhang et al. 2013; Lyu et al. 2014; Ma, Xie & Hou 2014; Wu et al. 2016; Liodakis et al. 2017; Zhu, Zhang & Fang 2019). Recently, Peng, Tang & Wang (2016) found that the tidal disruption event (TDE) has high-energy $\gamma$-ray emission (Swift J164449.3+573451, Swift J2058.4+0516, Swift J1112.2−8238). Curd & Narayan (2019) suggested that these TDEs with $\gamma$-ray emission have both a rapidly spinning black hole and magnetically arrested accretion (MAD) disc based on the general relativistic radiation magnetohydrodynamics (GRRMHD) simulations. The relativistic jets of these jetted TDEs can be powered by extracting the black hole rotation energy via an ordered magnetic field threading the ergosphere of a spinning black hole, namely the Blandford–Znajek mechanism (Dai, Lodato & Cheng 2021). Chen et al. (2021b) found that Fermi blazars can be explained by the MAD disc. Some authors suggested that the jets of blazars are likely governed by the Blandford–Znajek mechanism (e.g. Chai, Cao & Gu 2012; Zhang et al. 2015; Zhang, Liu & Fan 2022). These results may imply that the jet properties of the Fermi blazars are similar to those of TDEs. Some authors also found that the jets mechanism of GRBs may be the BZ mechanism (e.g. Lei et al. 2017; Xie, Lei & Wang 2017).

Since the successful launch of the *Fermi* telescope, many sources have detected high-energy $\gamma$-ray emissions, such as blazars, radio galaxies, and RLNLS1s (Abdo et al. 2009), which implies that these supermassive black hole with $\gamma$-ray emissions has strong relativistic jets. Previously, the properties of Fermi blazars have been studied based on a small sample. However, there have been questions such as what the relationship between Fermi blazars, radio galaxies, and RLNLS1 is? What is the jet formation mechanism of these jetted AGNs? Do they have similar jet properties? There has been a lack of research on this issue with a large sample. In this work, we use a large sample of $\gamma$-ray sources including blazars, radio galaxies, and RLNLS1s to study the properties of their jets. Section 2 presents the samples. Section 3 describes the results. Section 4 is the discussion, and Section 5 is the conclusions.

## 2 THE SAMPLE

### 2.1 The Fermi blazar and $\gamma$NLS1s sample

The Fermi Large Area Telescope (LAT) has released the fourth source catalogue data (4FGL-DR2; Abdollahi et al. 2020). First, Paliya et al. (2021) cross-matched the 4FGL-DR2 catalogue with the 16th data release of the Sloan Digital Sky Survey (SDSS-DR16; Ahumada et al. 2020)). Second, they searched the published optical spectrum of all remaining blazars in the literature using the NASA Extragalactic Database and SIMBAD Astronomical Database. Third, they searched the published black hole mass and accretion disc luminosity in the literature for objects leftover after completing the above steps. Finally, they got 1077 sources with reliable black hole mass and accretion disc luminosity.

### 2.2 Black hole mass and disc luminosity

Paliya et al. (2021) derived the black hole mass from the following three methods. First, the black hole mass is estimated by the virial method. The virial black hole mass can be calculated by using the following formula (Shen et al. 2011):

$$\log\left(\frac{M}{M_\odot}\right) = \alpha + \beta \log\left(\frac{\lambda L_\lambda}{10^{44} \text{erg s}^{-1}}\right) + 2\log\left(\frac{\text{FWHM}}{\text{km s}^{-1}}\right), \quad (1)$$

where $\lambda L_\lambda$ is the continuum luminosity. For H$\beta$, $\lambda L_\lambda$ is 5100 Å, 3000 Å (for Mg II), 1350 Å (for C IV). The $\alpha$ and $\beta$ are taken from McLure & Dunlop (2004) and Vestergaard & Peterson (2006). They also used the H$\alpha$ line to estimate the black hole mass. The formula is as follows:

$$\log\left(\frac{M_{\rm BH}}{M_\odot}\right) = 0.379 + 0.43 \log\left(\frac{L_{H\alpha}}{10^{42} erg \ s^{-1}}\right) + 2.1 \log\left(\frac{\text{FWHM}_{H\alpha}}{\text{km s}^{-1}}\right). \quad (2)$$

Second, they used the stellar velocity dispersion to estimate the black hole mass when these sources have no broad emission lines. The formula is as follows (Gültekin et al. 2009):

$$\log\left(\frac{M_{\rm BH}}{M_\odot}\right) = (8.12 \pm 0.08) + (4.24 \pm 0.41) \times \log\left(\frac{\sigma_*}{200 \text{km s}^{-1}}\right). \quad (3)$$

Third, the black hole mass is calculated by using the bulge luminosity. The formulas are as follows (Graham 2007):

$$\log\left(\frac{M_{\rm BH}}{M_\odot}\right) = \begin{cases} (-0.38 \pm 0.06)(M_R + 21) + (8.11 \pm 0.11), \\ (-0.38 \pm 0.06)(M_K + 24) + (8.26 \pm 0.11). \end{cases} \quad (4)$$

where $M_R$ and $M_K$ are the absolute magnitudes of the host galaxy bulge in the $R$ and $K$ bands, respectively.

We also note that the black hole mass is calculated using different methods in our sample. Tremaine et al. (2002) suggested that the uncertainty of black hole mass was calculated by using the stellar velocity dispersion is small, ≤0.25 dex. The uncertainty on the zero point of the line width luminosity–mass relation is approximately 0.5 dex (Gebhardt et al. 2000; Ferrarese et al. 2001). McLure & Dunlop (2001) suggested that the uncertainty of black hole mass was estimated by using the $M_{\rm BH}$–$M_R(M_H)$ relation is 0.6 dex.

The BLR luminosities given in Celotti, Padovani & Ghisellini (1997) were derived by scaling several strong emission lines to the quasar template spectrum of Francis et al. (1991), using Lya as a reference. We then assigned a reference value of 100 (hereafter the asterisk refers to luminosities in the same units) to Ly$\alpha$ emission and summed the line ratios (with respect to Ly$\alpha$) reported in Francis et al. (1991) and Celotti et al. (1997). This gives a total BLR fraction is $<L_{\rm BLR}> = 555.77 \sim 5.6$ Ly$\alpha$. The BLR luminosities are derived by using the following formula:

$$L_{BLR} = L_{line} \times \frac{<L_{BLR}>}{L_{rel.frac.}} \quad (5)$$

where $L_{line}$ is the emission line luminosity, and $L_{rel.frac.}$ is the line ratio, for H $\alpha$, H $\beta$, Mg II, and C IV are 77, 22, 34, and 63, respectively (Francis et al. 1991; Celotti et al. 1997). The accretion disc luminosity is estimated by using $L_{disc} = 10L_{BLR}$ (e.g. Baldwin & Netzer 1978), with an average uncertainty of a factor 2 (Calderone et al. 2013; Ghisellini et al. 2014).

### 2.3 Jet kinetic power

The jet kinetic power of Fermi sources is estimated by using the following formula (Cavagnolo et al. 2010):

$$\log P_{jet} = 0.75(\pm 0.14) \log P_{1.4} + 1.91(\pm 0.18). \quad (6)$$

The scatter of this relation is ≈0.78 dex. The $P_{1.4}$ is the radio luminosity which are estimated by using $P_{1.4} = 4\pi d_L^2 (1+z)^{\alpha-1} \nu S_\nu$, $S_\nu$ is the flux density, $z$ is the redshift, $d_L$ is the luminosity distance, $\alpha$ is radio spectral index, $\alpha = 0$ are adopted (Abdo et al. 2010; Komossa, Xu & Wagner 2018).

We carefully checked the sample of Paliya et al. (2021) and compared it with the source classification of Abdollahi et al. (2020) and Foschini et al. (2021), and found that 17 $\gamma$-ray narrow-line Seyfert 1 galaxies ($\gamma$NLS1s) were included in the sample of Paliya et al. (2021). We only consider these sources with 1.4 GHz radio flux from the NED. Finally, we get 504 FSRQs, 277 BL Lacs, and 17 $\gamma$NLS1s. The data are listed in Table 1.

### 2.4 The radio galaxies sample

We select the radio galaxies from the 4FGL-DR2 catalogue. The Fermi LAT has detected 41 radio galaxies (Ajello et al. 2020). We only consider the radio galaxies with reliable redshift, 1.4 GHz radio flux, and absolute magnitude ($M_H$). Finally, we get 39 radio galaxies. The 1.4 GHz radio flux comes from the NED. The black hole mass of radio galaxies is estimated by using the equation (4), namely $M_{BH}$–$M_R$. We get $M_H$ from the NED and use $R - H = 2.5$ to get the $M_R$ (Mannucci et al. 2001; Buttiglione et al. 2010). The disc luminosity of radio galaxies comes from the work of Buttiglione et al. (2010). We get seven radio galaxies with disc luminosity. We also use equation (6) to estimate the jet power of radio galaxies.

We also note that radio galaxies are divided into two types: FR I and FR II, adopting a threshold of $L_{1.4 GHz} = 10^{25}$ W Hz$^{-1}$ (Fanaroff & Riley 1974; Angioni 2020). The 1.4 GHz radio luminosity of radio galaxies with large radio luminosities comes from the lobe, not from the core, such as FR II radio galaxies. If these samples are going to be compared, it should be discussed why this is appropriate. Sikora, Stawarz & Lasota (2007) investigated how the total radio luminosity of AGN-powered radio sources depends on their accretion luminosity and the central black hole mass by using the sample of radio-loud broad-line AGNs [broad-line radio galaxies (BLRGs) plus radio-loud quasars], Seyfert galaxies and LINERs, FR I radio galaxies and optically selected quasars. They find that AGNs form two distinct and well-separated sequences on the radio-loudness-Eddington-ratio (or total 5GHz radio luminosity versus $B$-band nuclear luminosity) plane. The 'upper' sequence is formed by radio-selected AGNs, and the 'lower' sequence contains mainly optically selected quasars. They speculated that almost all BLRGs and radio-loud quasars in their samples have FR II radio morphology. They suggested that the difference can be explained by black hole mass and/or radio loudness. The 'upper' sequence has a large black hole mass ($M_{BH} > 10^8$ M$_\odot$), and the 'lower' sequence has a low black hole mass. We find that almost radio galaxies (FR I and FR II) in our sample have $M_{BH} > 10^8$ M$_\odot$. At the same time, Abdo et al. (2010) found that radio galaxies with $\gamma$-ray emission have high core dominance (CD) parameters. Paliya, Saikia & Stalin (2023) found that FR II radio galaxies (4FGL J1435.5+2021) have high core dominance parameters (log $CD = -0.11$), bright core and two hotspots, which is similar to other $\gamma$-ray detected FR I radio galaxies. Therefore, to sum up, it is appropriate when core and core plus lobe samples with $\gamma$-ray emission are going to be compared.

## 3 RESULTS

### 3.1 Intrinsic $\gamma$-ray luminosity

Nemmen et al. (2012) studied the relation between jet power and intrinsic $\gamma$-ray luminosity for Fermi blazars and $\gamma$-ray bursts (GRBs). They used the beaming factor ($f_b$) to correct the observation $\gamma$-ray luminosity ($L^{iso}$), $L^{int} = f_b L^{iso}$. For AGNs, we assume the jet luminosity $L^{int}$ approximately to be concentrated in a cone with a jet half-opening angle $\theta_j$, which is the product of isotropic luminosity $L^{iso}$ and observed beaming factor $f_b$. A conical jet will not light up the full celestial sphere but rather a fraction, the so-called beaming fraction $f_b$ (see details in Rhoads 1999; Frail et al. 2001). The $f_b$ was estimated by using $f_b = 1 - \cos(\theta_j)$, because jet opening angle $\theta_j \approx 1/\Gamma \ll 1$ (Jorstad et al. 2005; Pushkarev et al. 2009) and $\Gamma = (1-\beta^2)^{-1/2}$ ($\beta$ is the intrinsic velocity), thus $f_b = 1 - \cos(1/\Gamma)$, where $\Gamma$ is the bulk Lorentz factor of the flow. The Doppler-boosting factor is expressed as $\delta = 1/[\Gamma(1 - \beta \cos\theta_j)]$. If we assume that the Lorentz factor is equal to the Doppler factor ($\Gamma = \delta$, Ghisellini et al. 2010) and the continuous jet, we can obtain $f_b = 1/\delta^2$, and $L^{int} = L^{iso}/\delta^2$. Nemmen et al. (2012) got the bulk Lorentz factor from the works of Hovatta et al. (2009) and Pushkarev et al. (2009). They used the power-law fit of $f_b \approx 5 \times 10^{-4}(L_{49}^{iso})^{-0.39\pm0.15}$ to get other sources without beaming factor. However, they only got 41 Fermi blazars with bulk Lorentz factors. Liodakis et al. (2018) estimated the bulk Lorentz factor of 1029 sources observed by the Owens Valley Radio Observatory's 40 m telescope. We cross-match our sample with the work of Liodakis et al. (2018) and get 160 Fermi sources with bulk Lorentz factors. Following the Nemmen et al. (2012), we use $1 - \cos(1/\Gamma)$ to get $f_b$. Because $\Gamma$ is not available for the whole sample, we also use the power-law fit of $L^{iso}$ versus $f_b$ as an estimator for $f_b$. The $L^{iso}$ are calculated using $L^{iso} = 4\pi d_L^2 S_\gamma (1+z)^{\alpha_\gamma - 2}$. The uncertainty in $L^{iso}$ is calculated by using the propagating the error associated with $\alpha_\gamma$ and $S_\gamma$ quoted in the 4FGL. The average uncertainty in $L^{iso}$ corresponds to 0.06 dex. The average uncertainty in $f_b$ is 0.3 dex. The uncertainty of $L^{int}$ is 0.26 dex for the sources with the direct estimation of $f_b$ available (Nemmen et al. 2012). For the sources without direct estimation of $f_b$, we estimate the uncertainty in $L^{int}$ using the $L^{iso}$–$f_b$ relation, the average uncertainty is 0.44 dex. Therefore, we assume that the average uncertainty of $L^{int}$ in all sources is 0.35 dex. The relation between $L^{iso}$ and $f_b$ is shown in Fig. 1. We find a significant correlation coefficient $r = -0.46$ (significance level $P = 4.5 \times 10^{-10}$, significant correlation $P < 0.01$ confidence level).

### 3.2 Jet power versus intrinsic $\gamma$-ray luminosity

Fig. 2 shows jet power as a function of intrinsic $\gamma$-ray luminosity. We find a significant Pearson correlation between them for the whole sample ($r = 0.94$, $P < 0.0001$). Spearman and Kendall tau are used

**Table 1.** The sample of jetted AGNs.

| Name (1) | Type (2) | Redshift (3) | γ-ray energy flux (4) | Photo index (5) | log (M/M☉) (6) | log $L_{BLR}$ (7) | log $L^{iso}$ (8) | log $f_b$ (9) | log $L^{int}$ (10) | $f_{1.4\ GHz}$ (11) | log $P_{jet}$ (12) |
|---|---|---|---|---|---|---|---|---|---|---|---|
| J0001.5+2113 | FSRQ | 0.439 | 1.93E-11 | 2.680 | 7.539 | 43.65 | 46.54 | −2.331 | 44.21 | 0.217 | 46.10 |
| J0003.2+2207 | BLL | 0.1 | 1.73E-12 | 2.276 | 8.100 | 41.74 | 43.71 | −1.723 | 41.99 | 0.0087 | 43.88 |
| J0004.4−4737 | FSRQ | 0.88 | 6.51E-12 | 2.415 | 8.280 | 44.10 | 47.16 | −2.465 | 44.70 | 0.932 | 47.30 |
| J0006.3−0620 | BLL | 0.347 | 1.43E-12 | 2.171 | 8.924 | 43.52 | 45.01 | −1.341 | 43.67 | 2.051 | 46.62 |
| J0010.6+2043 | FSRQ | 0.598 | 1.95E-12 | 2.317 | 7.861 | 44.34 | 45.94 | −2.203 | 43.74 | 0.158 | 46.30 |
| J0011.4+0057 | FSRQ | 1.491 | 5.86E-12 | 2.320 | 8.664 | 44.71 | 48.20 | −2.688 | 45.51 | 0.167 | 47.46 |
| J0013.6+4051 | FSRQ | 0.256 | 2.08E-12 | 2.212 | 7.022 | 42.13 | 44.80 | −1.958 | 42.84 | 1.65 | 46.30 |
| J0013.6−0424 | FSRQ | 1.076 | 2.21E-12 | 2.359 | 7.816 | 44.03 | 47.07 | −2.445 | 44.62 | 0.304 | 47.19 |
| J0013.9−1854 | BLL | 0.095 | 3.15E-12 | 1.966 | 9.650 | 42.27 | 43.91 | −1.766 | 42.14 | 0.0295 | 44.24 |
| J0014.1+1910 | BLL | 0.477 | 2.30E-12 | 2.264 | 7.465 | 43.32 | 45.66 | −2.142 | 43.52 | 0.154 | 46.07 |
| J0014.2+0854 | BLL | 0.163 | 2.68E-12 | 2.498 | 8.850 | 42.37 | 44.42 | −1.876 | 42.54 | 0.326 | 45.42 |
| J0014.3−0500 | FSRQ | 0.791 | 5.09E-12 | 2.345 | 7.928 | 43.93 | 46.83 | −2.394 | 44.44 | 0.0318 | 46.08 |
| J0015.6+5551 | BLL | 0.217 | 4.02E-12 | 1.908 | 9.680 | 43.05 | 44.87 | −1.973 | 42.90 | 0.0849 | 45.20 |
| J0016.2−0016 | FSRQ | 1.577 | 9.95E-12 | 2.727 | 8.522 | 44.77 | 48.73 | −2.802 | 45.93 | 0.957 | 48.12 |
| J0016.5+1702 | FSRQ | 1.721 | 3.08E-12 | 2.631 | 8.874 | 44.74 | 48.41 | −2.733 | 45.68 | 0.135 | 47.62 |
| J0017.5−0514 | FSRQ | 0.227 | 1.22E-11 | 2.535 | 7.831 | 43.74 | 45.46 | −1.870 | 43.59 | 0.178 | 45.48 |
| J0017.8+1455 | BLL | 0.303 | 2.78E-12 | 2.179 | 8.270 | 43.16 | 45.13 | −2.028 | 43.10 | 0.0595 | 45.36 |
| J0019.6+7327 | FSRQ | 1.781 | 2.11E-11 | 2.594 | 9.306 | 45.62 | 49.32 | −1.967 | 47.35 | 1.25 | 48.40 |
| J0021.6−0855 | BLL | 0.648 | 1.86E-12 | 2.350 | 8.540 | 43.63 | 46.06 | −2.228 | 43.83 | 0.0472 | 45.99 |
| J0022.0+0006 | BLL | 0.306 | 1.91E-12 | 1.472 | 8.020 | 42.79 | 44.90 | −1.979 | 42.92 | 0.0042 | 44.50 |
| J0023.7+4457 | FSRQ | 1.062 | 5.84E-12 | 2.442 | 7.709 | 44.09 | 47.49 | −2.536 | 44.95 | 0.141 | 46.92 |
| J0024.7+0349 | FSRQ | 0.546 | 2.41E-12 | 2.389 | 7.114 | 43.62 | 45.91 | −2.196 | 43.71 | 0.022 | 45.57 |
| J0025.2−2231 | FSRQ | 0.834 | 1.48E-12 | 2.401 | 8.492 | 44.71 | 46.41 | −2.304 | 44.11 | 0.202 | 46.74 |
| J0028.4+2001 | FSRQ | 1.553 | 6.21E-12 | 2.436 | 8.426 | 44.57 | 48.37 | −2.725 | 45.65 | 0.287 | 47.70 |
| J0032.4−2849 | BLL | 0.324 | 2.35E-12 | 2.297 | 8.470 | 43.02 | 45.16 | −2.035 | 43.13 | 0.161 | 45.74 |
| J0038.2−2459 | FSRQ | 0.498 | 3.42E-12 | 2.364 | 8.139 | 43.97 | 45.92 | −2.198 | 43.72 | 0.413 | 46.43 |
| J0039.0−0946 | FSRQ | 2.106 | 4.64E-12 | 2.717 | 8.499 | 44.73 | 49.19 | −2.901 | 46.29 | 0.154 | 48.01 |
| J0040.4−2340 | BLL | 0.213 | 2.25E-12 | 2.126 | 8.680 | 42.75 | 44.62 | −1.919 | 42.70 | 0.0536 | 45.04 |
| J0040.9+3203 | FSRQ | 0.632 | 2.14E-12 | 2.407 | 7.173 | 43.47 | 46.09 | −2.235 | 43.86 | 0.414 | 46.67 |
| J0042.2+2319 | FSRQ | 1.425 | 5.08E-12 | 2.289 | 8.733 | 44.49 | 48.02 | −2.649 | 45.37 | 1.27 | 48.05 |
| J0043.8+3425 | FSRQ | 0.969 | 2.62E-11 | 1.942 | 7.828 | 43.76 | 47.81 | −2.604 | 45.21 | 0.0933 | 46.67 |
| J0044.2−8424 | FSRQ | 1.032 | 3.82E-12 | 2.615 | 8.519 | 44.87 | 47.30 | −2.495 | 44.81 | 0.53 | 47.32 |
| J0045.1−3706 | FSRQ | 1.015 | 6.94E-12 | 2.561 | 8.608 | 44.86 | 47.51 | −2.540 | 44.97 | 0.33 | 47.14 |
| J0045.7+1217 | BLL | 0.255 | 1.06E-11 | 1.998 | 8.820 | 43.18 | 45.49 | −2.106 | 43.38 | 0.104 | 45.40 |
| J0047.9+2233 | FSRQ | 1.163 | 7.20E-12 | 2.527 | 8.072 | 44.57 | 47.80 | −2.602 | 45.20 | 0.0837 | 46.88 |
| J0049.0+2252 | BLL | 0.264 | 2.27E-12 | 2.240 | 9.040 | 42.63 | 44.88 | −1.975 | 42.91 | 0.076 | 45.32 |
| J0049.6−4500 | FSRQ | 0.121 | 2.23E-12 | 2.494 | 8.084 | 42.51 | 44.03 | −1.792 | 42.24 | 0.206 | 45.05 |
| J0050.0−5736 | FSRQ | 1.797 | 5.53E-12 | 2.603 | 9.064 | 45.88 | 48.77 | −2.811 | 45.96 | 2.11 | 48.59 |
| J0051.1−0648 | FSRQ | 1.975 | 9.21E-12 | 2.334 | 9.310 | 46.11 | 49.12 | −2.854 | 46.27 | 0.904 | 48.47 |
| J0056.3−0935 | BLL | 0.103 | 7.99E-12 | 1.871 | 8.960 | 42.22 | 44.39 | −1.869 | 42.52 | 0.201 | 44.93 |
| J0058.0−0539 | FSRQ | 1.246 | 6.11E-12 | 2.460 | 8.699 | 45.28 | 47.86 | −2.615 | 45.24 | 0.742 | 47.68 |
| J0058.4+3315 | FSRQ | 1.371 | 3.06E-12 | 2.361 | 8.577 | 44.26 | 47.74 | −2.589 | 45.15 | 0.154 | 47.31 |
| J0059.2+0006 | FSRQ | 0.719 | 2.38E-12 | 2.358 | 8.564 | 45.08 | 46.34 | −2.288 | 44.05 | 2.43 | 47.38 |
| J0059.3−0152 | BLL | 0.144 | 2.56E-12 | 1.784 | 8.630 | 42.52 | 44.23 | −1.835 | 42.39 | 0.018 | 44.38 |
| J0102.4+4214 | NLSY1 | 0.876 | 7.91E-12 | 2.700 | 8.329 | 44.66 | 47.31 | −2.497 | 44.81 | 0.0431 | 46.30 |
| J0102.8+5824 | FSRQ | 0.644 | 4.73E-11 | 2.288 | 9.005 | 45.04 | 47.44 | −2.423 | 45.02 | 0.849 | 46.92 |
| J0103.5+1526 | BLL | 0.246 | 2.34E-12 | 2.413 | 9.020 | 42.78 | 44.83 | −1.964 | 42.87 | 0.226 | 45.62 |
| J0103.8+1321 | BLL | 0.49 | 3.54E-12 | 2.167 | 9.690 | 43.41 | 45.87 | −2.187 | 43.68 | 0.0519 | 45.74 |
| J0104.8−2416 | FSRQ | 1.747 | 6.50E-12 | 2.616 | 8.980 | 45.05 | 48.77 | −2.811 | 45.96 | 0.235 | 47.82 |
| J0105.1+3929 | BLL | 0.44 | 6.63E-12 | 2.292 | 8.172 | 43.34 | 46.01 | −2.218 | 43.79 | 0.0915 | 45.82 |
| J0108.1−0039 | FSRQ | 1.375 | 3.89E-12 | 2.669 | 9.007 | 45.24 | 47.96 | −2.637 | 45.32 | 0.93 | 47.90 |
| J0108.6+0134 | FSRQ | 2.099 | 1.16E-10 | 2.353 | 9.632 | 45.62 | 50.41 | −3.189 | 47.22 | 2.62 | 48.94 |
| J0109.7+6133 | FSRQ | 0.783 | 3.53E-11 | 2.607 | 7.480 | 42.39 | 47.72 | −2.928 | 44.79 | 0.305 | 46.80 |
| J0111.4+0534 | BLL | 0.347 | 1.63E-12 | 1.952 | 8.470 | 43.06 | 45.04 | −2.009 | 43.03 | 0.0165 | 45.05 |

*Notes.* Columns (1) is the name of sources; Columns (2) is the Type of sources; Columns (3) is redshift; Columns (4) is the energy flux of γ-ray in units erg cm$^{-2}$ s$^{-1}$; Columns (5) is the photo index ($\alpha_\gamma$); Columns (6) is the black hole mass; Columns (7) is the BLR luminosity, units is erg s$^{-1}$; Columns (8) is the observation γ-ray luminosity, units is erg s$^{-1}$; Columns (9) is the beaming factor; Columns (10) is the intrinsic γ-ray luminosity; Columns (11) is the 1.4 GHz radio flux in units jy; Columns (12) is the jet power in units erg s$^{-1}$. This table is available in its entirety in machine-readable form.

to detect this correlation. The Spearman correlation coefficient and significance level are $r = 0.94$ and $P < 0.0001$. The Kendall tau correlation coefficient and significance level are $r = 0.79$ and $P = 1.91 \times 10^{-255}$. These two tests also show a significant correlation.

Partial regression analysis also shows that the linear correlation between intrinsic γ-ray luminosity and jet power is significantly correlated when the effects of redshift are removed ($r_{XY,z} = 0.69$, $P = 4.54 \times 10^{-117}$). It can be seen that the jet power and the intrinsic

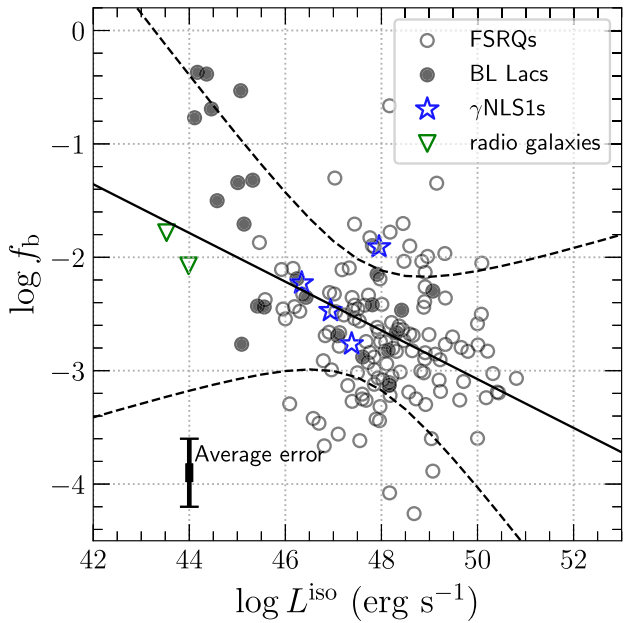
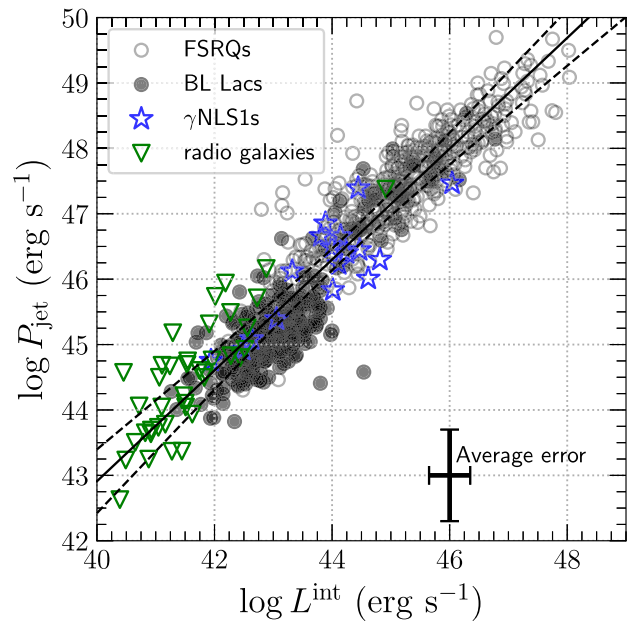

**Figure 1.** The relation between observation $\gamma$-ray luminosity ($L^{\rm iso}$) and beaming factor for 163 Fermi sources. The solid lines correspond to the best-fitting linear models obtained with the symmetric least-squares fit and are given by $\log f_b = (-0.21 \pm 0.03) \log L^{\rm iso} + (7.67 \pm 1.54)$. The dashed lines indicate the $3\sigma$ confidence band. The black empty dot is for FSRQs. The black dot is for BL Lacs. The blue star is for $\gamma$NLS1s. The upside-down triangle is for radio galaxies.

**Figure 2.** The relation between intrinsic $\gamma$-ray luminosity ($L^{\rm int}$) and jet power for the whole sample. The solid line corresponds to the best-fitting linear models obtained with the symmetric least-squares fit. The dashed lines indicate $3\sigma$ confidence bands.

$\gamma$-ray luminosity are correlated and can be well-fitted with the form $P_{\rm jet} \propto L^p_{\rm int}$. The fitting formula is

$$\log P_{\rm jet} = (0.85 \pm 0.01) \log L^{\rm int} + (9.01 \pm 0.48) \quad (7)$$

for the whole sample,

$$\log P_{\rm jet} = (0.70 \pm 0.02) \log L^{\rm int} + (15.88 \pm 0.75) \quad (8)$$

for FSRQs ($r = 0.88$, $P = 1.18 \times 10^{-167}$)

$$\log P_{\rm jet} = (0.83 \pm 0.03) \log L^{\rm int} + (9.52 \pm 1.23) \quad (9)$$

for BL Lacs ($r = 0.87$, $P = 5.38 \times 10^{-86}$),

$$\log P_{\rm jet} = (0.68 \pm 0.11) \log L^{\rm int} + (16.49 \pm 4.88) \quad (10)$$

for $\gamma$NLS1s ($r = 0.84$, $P = 2.11 \times 10^{-5}$),

$$\log P_{\rm jet} = (0.93 \pm 0.09) \log L^{\rm int} + (5.45 \pm 4.07) \quad (11)$$

for radio galaxies ($r = 0.84$, $P = 1.42 \times 10^{-11}$). The Spearman and Kendall tau test also shows a significant correlation (Table 2). We find that the $L^{\rm int}$–$P_{\rm jet}$ relations of FSRQs and BL Lacs are consistent with the $L^{\rm int}$–$P_{\rm jet}$ of $\gamma$NLS1s and radio galaxies. It implies that there exists a universal correlation between $\gamma$-ray luminosity and jet power among all the relativistic jets. In other words, once 'black hole engines' produce relativistic jets, they seem to do so maintaining the same coupling between the total power carried by jet and power radiated away. This universal scaling for the energetics of jets is maintained across the mass scale, regardless of the different environments and accretion flow conditions around the compact object.

### 3.3 Black hole mass versus intrinsic $\gamma$-ray luminosity

The relation between intrinsic $\gamma$-ray luminosity and black hole mass is shown in Fig. 3. Pearson analysis is applied to analyse the correlation between black hole mass and intrinsic $\gamma$-ray luminosity for all sources. We find that there is a significant correlation between black hole mass and intrinsic $\gamma$-ray luminosity for the whole sample ($r = 0.14$, $P = 5.86 \times 10^{-5}$). The test of Spearman ($r = 0.13$, $P < 0.0001$) and Kendall tau ($r = 0.10$, $P = 8.26 \times 10^{-6}$) also show a significant correlation for the whole sample. Partial regression analysis also shows that the linear correlation between intrinsic $\gamma$-ray luminosity and black hole mass is significantly correlated when the effects of redshift are removed ($r_{XY, z} = -0.14$, $P < 0.0001$). These results suggest that jet power depends on the black hole mass.

From Fig. 3, we also find that some BL Lacs and $\gamma$NLS1s follow the relationship between the mass of the black hole and the intrinsic $\gamma$-ray luminosity in FSRQs. According to synchrotron peak frequency ($\nu_p$), we find that these BL Lacs are low peak frequency BL Lacs (LBLs, $\log \nu_p < 14.0$). Li et al. (2010) found that FSRQs and LBLs occupy the same region in $\alpha_{\rm ox}$–$\alpha_{x\gamma}$ plane, which suggests that they have similar spectral properties. Cha et al. (2014) suggested that the evolutionary track of Fermi blazars is from FSRQs to LBLs. Chen et al. (2015b) found that the accretion discs of FSRQs and LBLs are dominated by radiation pressure. Chen et al. (2021a) found that FSRQs and LBLs have the same particle acceleration mechanism.

Soares & Nemmen (2020) used 154 Fermi FSRQs to study the relation between black hole mass and $\gamma$-ray luminosity. They obtained $\log M \propto L^{0.37 \pm 0.05}$ for FSRQs. We get the slope of intrinsic $\gamma$-ray luminosity and black hole mass relation is $\log M \sim (0.29 \pm 0.02) \log L^{\rm int}$ for 504 FSRQs. Our slope is slightly smaller than theirs. The possible reasons are that our sample is larger than theirs and that we use the beaming factor to correct the $\gamma$-ray luminosity.

### 3.4 Intrinsic $\gamma$-ray luminosity versus disc luminosity

Intrinsic $\gamma$-ray luminosity is a good indicator of jet power. Thus, we studied the relation between intrinsic $\gamma$-ray luminosity and accretion

**Table 2.** The results of correlation analysis for sample.

| Sample | Pearson | | | | Spearman | | Kendall tau | |
|---|---|---|---|---|---|---|---|---|
| | A | B | r | p | r | p | r | p |
| $x = \log L^{\text{int}}; y = \log P_{\text{jet}}$ | | | | | | | | |
| Whole sample | $0.85 \pm 0.01$ | $9.01 \pm 0.48$ | 0.94 | <0.0001 | 0.94 | <0.0001 | 0.79 | $1.91 \times 10^{-255}$ |
| FSRQs | $0.70 \pm 0.02$ | $15.88 \pm 0.75$ | 0.88 | $1.18 \times 10^{-167}$ | 0.88 | $2.94 \times 10^{-165}$ | 0.71 | $1.62 \times 10^{-124}$ |
| BL Lacs | $0.83 \pm 0.03$ | $9.52 \pm 1.23$ | 0.87 | $5.38 \times 10^{-86}$ | 0.75 | $2.81 \times 10^{-51}$ | 0.58 | $4.73 \times 10^{-46}$ |
| $\gamma$-NLS1s | $0.68 \pm 0.11$ | $16.49 \pm 4.88$ | 0.84 | $2.11 \times 10^{-5}$ | 0.62 | 0.008 | 0.49 | 0.006 |
| Radio galaxies | $0.93 \pm 0.09$ | $5.45 \pm 4.07$ | 0.84 | $1.42 \times 10^{-11}$ | 0.81 | $2.69 \times 10^{-10}$ | 0.65 | $6.85 \times 10^{-9}$ |
| $x = \log L_{\text{disc}}; y = \log L^{\text{int}}$ | | | | | | | | |
| Whole sample | $1.05 \pm 0.02$ | $-3.10 \pm 1.08$ | 0.84 | $1.13 \times 10^{-214}$ | 0.84 | $3.63 \times 10^{-212}$ | 0.66 | $1.69 \times 10^{-171}$ |
| FSRQs | $0.94 \pm 0.05$ | $2.38 \pm 2.41$ | 0.62 | $6.04 \times 10^{-55}$ | 0.59 | $6.45 \times 10^{-49}$ | 0.42 | $5.03 \times 10^{-46}$ |
| BL Lacs | $1.14 \pm 0.05$ | $-7.26 \pm 2.14$ | 0.82 | $4.06 \times 10^{-68}$ | 0.82 | $5.97 \times 10^{-70}$ | 0.67 | $5.01 \times 10^{-62}$ |
| $\gamma$-NLS1s | $0.92 \pm 0.18$ | $2.37 \pm 7.98$ | 0.80 | 0.0001 | 0.75 | 0.0006 | 0.60 | 0.0004 |
| $x = \log L_{\text{disc}}; y = \log P_{\text{jet}}$ | | | | | | | | |
| Whole sample | $1.00 \pm 0.02$ | $1.44 \pm 0.89$ | 0.87 | $2.94 \times 10^{-251}$ | 0.87 | $7.26 \times 10^{-243}$ | 0.68 | $2.54 \times 10^{-182}$ |
| FSRQs | $0.83 \pm 0.04$ | $9.35 \pm 1.74$ | 0.70 | $4.82 \times 10^{-75}$ | 0.70 | $3.24 \times 10^{-76}$ | 0.52 | $2.93 \times 10^{-67}$ |
| BL Lacs | $1.00 \pm 0.05$ | $1.25 \pm 2.35$ | 0.75 | $3.98 \times 10^{-51}$ | 0.70 | $3.36 \times 10^{-42}$ | 0.52 | $1.03 \times 10^{-37}$ |
| $\gamma$-NLS1s | $0.73 \pm 0.15$ | $13.52 \pm 6.61$ | 0.79 | 0.0002 | 0.76 | 0.0004 | 0.62 | 0.0003 |

*Note.* The A is slope; B is the intercept; r is correlation coefficient; p is significance level ($p < 0.01$).

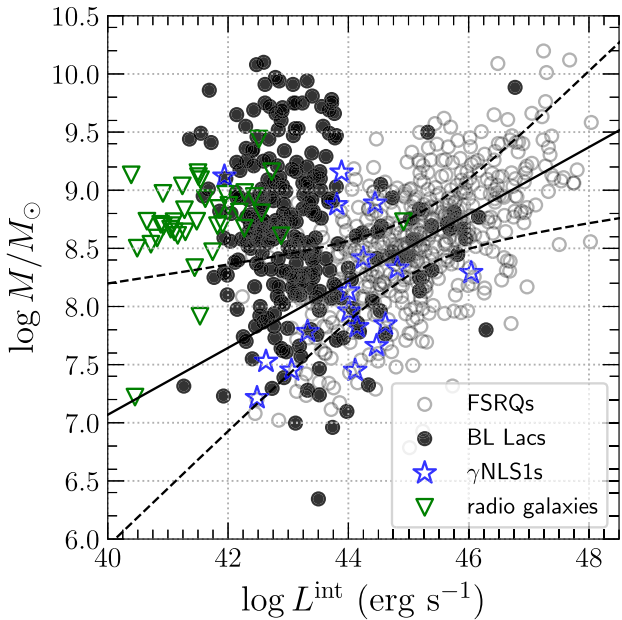

**Figure 3.** The relation between intrinsic $\gamma$-ray luminosity ($L^{\text{int}}$) and black hole mass for the whole sample. The solid line corresponds to the best-fitting linear models obtained with the symmetric least-squares fit for FSRQs and is given by $\log M = (0.29 \pm 0.02) \log L^{\text{int}} - (4.44 \pm 0.80)$. The dashed lines indicate $3\sigma$ confidence bands.

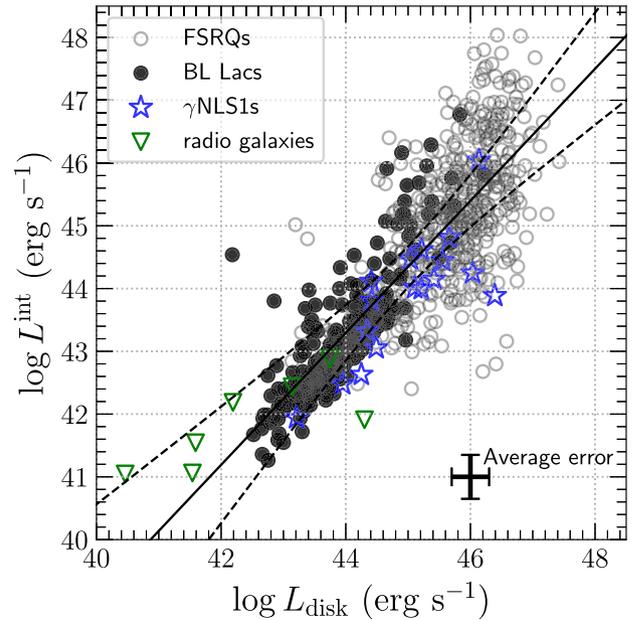

**Figure 4.** The relation between intrinsic $\gamma$-ray luminosity ($L^{\text{int}}$) and disc luminosity for the whole sample. The solid line corresponds to the best-fitting linear models obtained with the symmetric least-squares fit. The dashed lines indicate $3\sigma$ confidence bands.

disc luminosity. Fig. 4 shows the relationship between intrinsic $\gamma$-ray luminosity and disc luminosity for the whole sample. The Pearson analysis shows a significant correlation between intrinsic $\gamma$-ray luminosity and disc luminosity for the whole sample (see Table 2). The test of Spearman and Kendall tau also shows a significant correlation for the whole sample. Partial regression analysis also shows that the linear correlation between intrinsic $\gamma$-ray luminosity and disc luminosity is significant even after the effects of redshift are removed ($r_{XY,z} = 0.56$, $P = 2.47 \times 10^{-66}$). We also use correlation analysis for every single type of sample (see Table 2). The fitting formulas are

$$\log L^{\text{int}} = (1.05 \pm 0.02) \log L_{\text{disc}} + (-3.10 \pm 1.08) \quad (12)$$

for the whole sample,

$$\log L^{\text{int}} = (0.94 \pm 0.05) \log L_{\text{disc}} + (2.38 \pm 2.41) \quad (13)$$

for FSRQs,

$$\log L^{\text{int}} = (1.14 \pm 0.05) \log L_{\text{disc}} + (-7.26 \pm 2.14) \quad (14)$$

for BL Lacs,

$$\log L^{\text{int}} = (0.92 \pm 0.18) \log L_{\text{disc}} + (2.37 \pm 7.98) \quad (15)$$

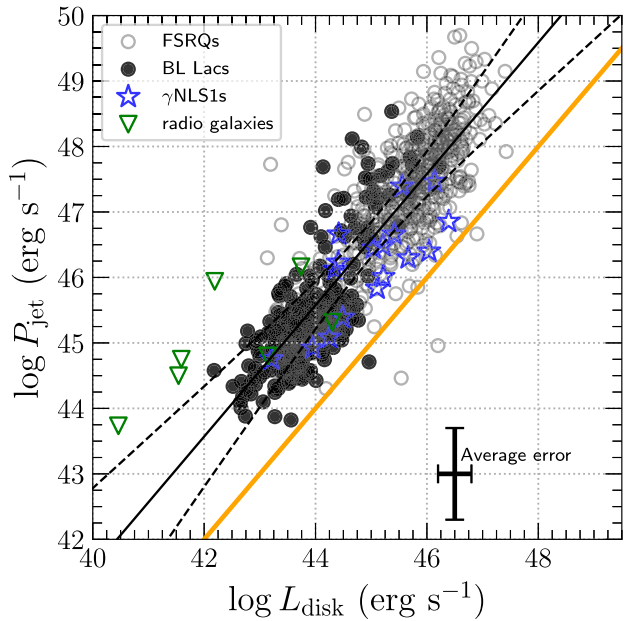

**Figure 5.** The relation between jet power and disc luminosity for the whole sample. The solid line corresponds to the best-fitting linear models obtained with the symmetric least-squares fit and is given by $\log P_{\rm jet} = (1.00 \pm 0.02)\log L_{\rm disc} + (1.44 \pm 0.90)$. The dashed lines indicate the $3\sigma$ confidence band. The orange solid line indicates $\log P_{\rm jet} = \log L_{\rm disc}$.

for $\gamma$NLS1s. Because only seven radio galaxies have accretion disc luminosity, we do not analyse the relationship between the intrinsic $\gamma$-ray luminosity and the accretion disc luminosity for the radio galaxies. From the above results, the slope indices of different subsamples are coherent with a value of $\sim 1$ within the error ranges.

### 3.5 Jet power versus disc luminosity

Fig. 5 shows the relation between jet power and disc luminosity for the whole sample. The Pearson analysis shows a significant correlation between jet power and disc luminosity for the whole sample (Table 2). The tests of Spearman and Kendall tau also show a significant correlation for the whole sample. We also obtain $\log P_{\rm jet} \sim (1.00 \pm 0.02)\log L_{\rm disc}$. Partial regression analysis also shows that the linear correlation between jet power and disc luminosity is significant when the effects of redshift are removed ($r_{\rm XY, z} = 0.67, P = 2.83 \times 10^{-104}$). At the same time, there is also a significant relation between jet power and disc luminosity for other types of AGNs (Table 2). These results suggest that the relationship between jet and accretion may be the same in various black hole systems. Merloni et al. (2003) found that the stellar mass black holes in XRBs and supermassive black holes follows the 'Fundamental Plane', $\log L_{\rm R} = 0.60\log L_{\rm X} + 0.78\log M + 7.33$. Falcke et al. (2004) suggested that hard and quiescent state XRBs, LINERs, FR I radio galaxies, and BL Lacs can be unified and fall on a common radio/X-ray correlation. Some authors also found that both the XRBs in their low/hard and LLAGNs can be successfully explained by the advection-dominated accretion flows (ADAFs; also called radiatively inefficient accretion flows) (e.g. Quataert et al. 1999; Yuan, Cui & Narayan 2005; Remillard & McClintock 2006; Wu & Cao 2008; Wu et al. 2013; Nemmen, Storchi-Bergmann & Eracleous 2014). Recently, Arcodia et al. (2020) compared the luminosities of the accretion disc and the corona in luminous AGNs to those of the prototypical XRBs GX 339-4 during its evolution through the soft state. They found a similar scatter in the disc and the corona luminosity distributions when the accretion rate and the (X-ray) power-law distributions are homogenized for both samples, suggesting that the (mass-scaled) common black hole accretion scheme might also hold during the soft state. Fernández-Ontiveros & Muñoz-Darias (2021) found that the accretion state in AGNs is similar to that of XRBs based on luminosity-excitation diagram (LED). These results may imply that the phenomenology of how black holes accrete matter is indeed somewhat analogous between AGNs and XRBs. Fender & Belloni (2004) concluded that the physics of disc–jet coupling in XRBs and AGNs are very closely linked and that by studying the nearby XRBs (GRS 1915+105). Foschini (2012) also found that jets in AGNs and XRBs are similarities.

From Fig. 5, we also find that the jet power of almost all jetted AGNs is larger than the luminosity of the accretion disc. According to the spectral energy distributions modelling, Ghisellini et al. (2014) also found that the jet power is larger than the accretion disc luminosity for 234 Fermi blazars. We confirm their conclusion. This jet power is somewhat larger than the luminosity of the accretion disc. There may be two explanations. One, this is not a coincidence, but the result of the catalysis of the magnetic field magnified by the disc. When the magnetic energy density exceeds the energy density ($\sim \rho c^2$) of the accreted material near the last stable orbit, the accretion stops and the magnetic energy decreases, as shown in numerical simulations (Tchekhovskoy, Narayan & McKinney 2011; Tchekhovskoy et al. 2014), which is confirmed by recent observational evidence (Zamaninasab et al. 2014). The other is similar to the jet and accretion coupling of XRBs. The ability of black hole systems to produce jets depends on the state of the accretion flow (Davis & Tchekhovskoy 2020). Fender, Gallo & Jonker (2003) proved that when the mass accretion rate is relatively low, the black hole XRBs should enter a 'jet-dominated' state, in which the majority of the liberated accretion power is in the form of a (radiatively inefficient) jet. Malzac, Merloni & Fabian (2004) studied the jet–disc coupling in the black hole XTE J1118+480. They suggested that the jet probably dominates the energetic output of all accreting black holes in the low/hard state. Falcke et al. (2004) suggested that the jet emission dominates the emission from the accretion flow for such a sub-Eddington state including X-ray binaries in the hard and quiescent states, the Galactic Centre (Sgr A*), FR I radio galaxies, a large fraction of BL Lac objects, and LLAGNs (e.g. Yuan et al. 2002).

### 3.6 Jet power versus broad-line region luminosity and Eddington luminosity

We use multiple linear regression analysis to obtain the relation between jet power and both BLR luminosity and Eddington luminosity for the whole sample with a 99 per cent confidence level and $r = 0.87$ (Fig. 6):

$$\log P_{\rm jet} = 1.01(\pm 0.02)\log L_{\rm BLR} - 0.05(\pm 0.04)\log L_{\rm Edd} \\ + 4.35(\pm 1.85), \quad (16)$$

for FSRQs ($r = 0.69, P = 5.72 \times 10^{-74}$),

$$\log P_{\rm jet} = 0.62(\pm 0.06)\log L_{\rm BLR} + 0.39(\pm 0.08)\log L_{\rm Edd} \\ + 1.90(\pm 2.35), \quad (17)$$

for BL Lacs ($r = 0.75, P = 1.17 \times 10^{-51}$),

$$\log P_{\rm jet} = 0.99(\pm 0.05)\log L_{\rm BLR} - 0.10(\pm 0.05)\log L_{\rm Edd} \\ + 7.03(\pm 3.45), \quad (18)$$

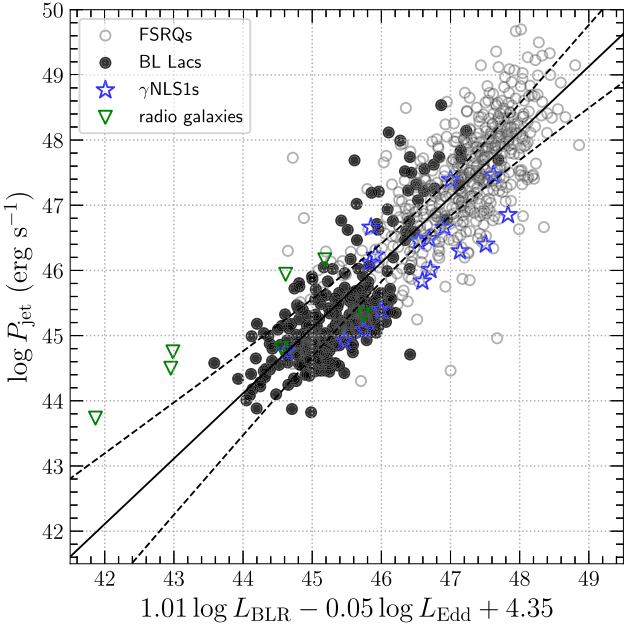
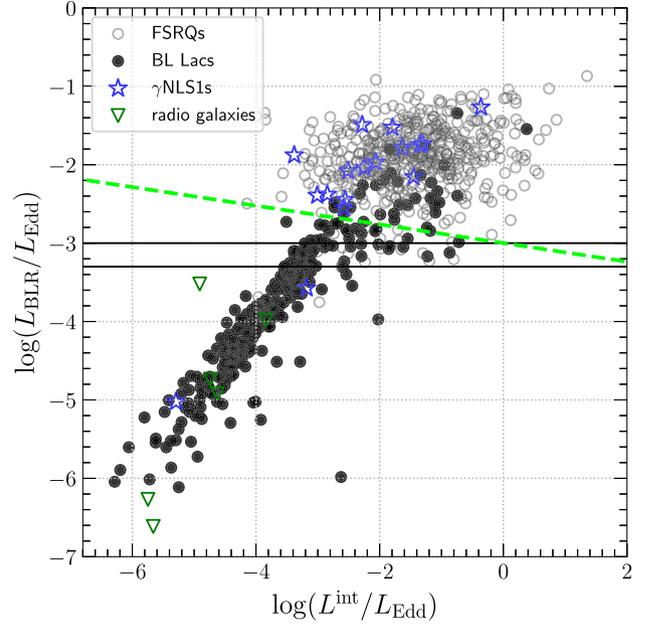

**Figure 6.** The jet power as a function of both BLR luminosity and Eddington luminosity for the whole sample. The solid line corresponds to the best-fitting linear models obtained with the symmetric least-squares fit. The dashed lines indicate 3σ confidence band.

**Figure 7.** Broad-line luminosity as a function of intrinsic γ-ray luminosity both in Eddington units for the whole sample. The horizontal solid lines indicate the luminosity divide between FSRQs and BL Lacs at $L_{BLR}/L_{Edd} \sim 10^{-3}$ and $L_{BLR}/L_{Edd} \sim 5 \times 10^{-4}$ from Sbarrato et al.(2012) and Ghisellini et al. (2011). The green dashed dividing line is our best result from the support vector machine.

for $\gamma$NLS1s ($r = 0.78$, $P = 0.0002$),

$$\log P_{jet} = 0.68(\pm 0.15) \log L_{BLR} + 0.23(\pm 0.21) \log L_{Edd} + 5.69(\pm 10.17), \quad (19)$$

Using $L_{bol} \approx 10 L_{BLR}$ (Netzer 1990), equations (16)–(19) can be expressed in a different form as

$$\log P_{jet} = 1.01 \log L_{bol}/L_{Edd} + 0.96 \log M + 39.93 \quad (20)$$

for the whole sample,

$$\log P_{jet} = 0.62 \log L_{bol}/L_{Edd} + 1.01 \log M + 39.77 \quad (21)$$

for FSRQs,

$$\log P_{jet} = 0.99 \log L_{bol}/L_{Edd} + 0.89 \log M + 39.96 \quad (22)$$

for BL Lacs,

$$\log P_{jet} = 0.68 \log L_{bol}/L_{Edd} + 0.91 \log M + 39.69 \quad (23)$$

for $\gamma$NLS1s. Theoretically, Heinz & Sunyaev (2003) suggested that the jet power depends on the black hole mass and accretion rate in core-dominated jets: for standard accretion, $F_\nu \sim M^{17/12}$; for radiatively inefficient accretion modes, $F_\nu \sim (\dot{m}M)^{17/12}$. Many observations have confirmed this theory. Merloni et al. (2003) and Falcke et al. (2004) proposed the fundamentals of black holes. The radio luminosity depends on the black hole mass and X-ray luminosity. Foschini (2014) studied the unification of relativistic jets from compact objects. They found that the existence of a secondary branch in AGN is similar to what was already known in Galactic binaries. Foschini (2014) suggested that the jet power can be scaled as $\log P_{jet} \propto \frac{17}{12} \log M$ for radiation-pressure-dominated accretion disc; $\log P_{jet} \propto \frac{17}{12} \log M + \frac{1}{2} \log \frac{L_{disc}}{L_{Edd}}$ for gas-pressure-dominated accretion disc. Wang, Luo & Ho (2004) studied the properties of relativistic jets of AGN and obtained $\log P_{jet} = 0.25(\pm 0.09) \log L_{BLR} + 0.65(\pm 0.25) \log L_{Edd} + 5.07(\pm 10.05)$. According to studying the jet power, radio loudness, and black hole mass in radio-loud AGNs, Liu, Jiang & Gu (2006) found $\log P_{jet} = 0.22 \log L_{bol}/L_{Edd} + 0.59 \log M + 40.48$. Foschini (2012) found that the relation between jet power, black hole mass, and the disc luminosity in Eddington units for AGNs and XRBs is $\log P_{jet} = (0.44 \pm 0.03) \log \frac{L_{disc}}{L_{Edd}} + 1.4 \log M + (34.70 \pm 0.07)$.

We compare our results with these results from other authors and find that our results are similar to results from other authors, i.e. the dependence of jet power on both the Eddington ratio and black hole mass. We define the contribution rates of the Eddington ratio and black hole mass to the jet power as $\epsilon_{Eddingtonratio} = 1.01/(1.01 + 0.96) \times 100$ per cent = 51 per cent and $\epsilon_{mass} = 0.96/(1.01 + 0.96) \times 100$ per cent = 49 per cent for the whole sample; $\epsilon_{Eddingtonratio} = 0.62/(0.62 + 1.01) \times 100$ per cent = 38 per cent and $\epsilon_{mass} = 1.01/(0.62 + 1.01) \times 100$ per cent = 62 per cent for FSRQs, $\epsilon_{Eddingtonratio} = 0.99/(0.99 + 0.89) \times 100$ per cent = 52.7 per cent and $\epsilon_{mass} = 0.89/(0.99 + 0.89) \times 100$ per cent = 47.3 per cent for BL Lacs, $\epsilon_{Eddingtonratio} = 0.68/(0.68 + 0.91) \times 100$ per cent = 42.8 per cent and $\epsilon_{mass} = 0.91/(0.68 + 0.91) \times 100$ per cent = 57.2 per cent for $\gamma$NLS1s. We find that the contribution rate of black hole mass to jet power is greater than that of Eddington ratio to jet power for FSRQs and $\gamma$NLS1s. However, the contribution rate of the Eddington ratio to jet power is greater than that of black hole mass to jet power for BL Lacs.

### 3.7 Divide between BL Lacs and FSRQs

Ghisellini et al. (2011) and Sbarrato et al. (2012) proposed a physical distinction between FSRQs and BL Lacs by using the $L_{BLR}/L_{Edd}$ versus $L_\gamma/L_{Edd}$. Ghisellini et al. (2011) got that the FSRQs and BL Lacs can be divided by using $L_{BLR}/L_{Edd} \sim 5 \times 10^{-4}$. Sbarrato et al. (2012) used $L_{BLR}/L_{Edd} \sim 10^{-3}$ to divide FSRQs and BL Lacs. Fig. 7 shows the relation between $L_{BLR}/L_{Edd}$ and $L^{int}/L_{Edd}$ for the whole sample. The FSRQs and BL Lacs in our sample can be roughly separated according to the divide lines of Ghisellini et al. (2011) and Sbarrato et al. (2012). To better separate the FSRQs and BL Lacs, we use a support vector machine (SVM), a kind of machine-learning method, to redivide the FSRQs and BL Lacs. The result of our dividing line gives an accuracy of 95 per cent and the dividing line is as follows:

$$\log \frac{L_{\rm BLR}}{L_{\rm Edd}} = -0.12 \log \frac{L^{\rm int}}{L_{\rm Edd}} - 2.99 \tag{24}$$

From Fig. 7, we also find that most of $\gamma$NLS1s are in the region occupied by FSRQs, while radio galaxies are in the region occupied by BL Lacs. These results imply that the physical properties of $\gamma$NLS1s are similar to those of FSRQs, namely radiatively efficient accretion disc. The physical properties of radio galaxies are similar to those of BL Lacs, namely radiatively inefficient accretion disc. These radio galaxies have $\log L^{\rm int}/L_{\rm Edd} < -3$. Abdo et al. (2010) suggested that the sources with $\log L_{\gamma}/L_{\rm Edd} \leq -2$ may be classified as radio galaxies. Our results confirm their conclusion.

## 4 DISCUSSION

### 4.1 Intrinsic $\gamma$-ray luminosity

We use the beaming factor ($f_{\rm b}$) to correct the observed $\gamma$-ray luminosity and obtain the intrinsic $\gamma$-ray luminosity. There is a significant correlation between the intrinsic $\gamma$-ray luminosity and jet power for our sample. We derive the $\log P_{\rm jet} \sim (0.85 \pm 0.01) \log L^{\rm int}$ for the whole sample. Zhang et al. (2013) used the SED model to obtain the jet power and synchrotron peak luminosity. They studied the relationship between jet power and intrinsic synchrotron peak luminosity and derived $P_{\rm jet} \propto L_{\rm s,jet}^{0.79\pm0.01}$ for blazars and $\gamma$-ray bursts. Our results are consistent with theirs. At the same time, we also find that the slope of $\log P_{\rm jet}$–$\log L^{\rm int}$ of different subsamples is the same within the error range. Zhu et al. (2019) found a relation between jet power and intrinsic $\gamma$-ray luminosity ($L_{\rm jet}$) for a short $\gamma$-ray burst (SGRBs): $\log P_{\rm jet}^{\rm SGRBs} = (0.84 \pm 0.15) \log L_{\rm jet}^{\rm SGRBs} + (7.72 \pm 7.57)$; for long $\gamma$-ray bursts (LGRBs): $\log P_{\rm jet}^{\rm LGRBs} = (0.82 \pm 0.07) \log L_{\rm jet}^{\rm LGRBs} + (9.45 \pm 3.62)$. Ma et al. (2014) found that the relation jet power and intrinsic $\gamma$-ray luminosity for XRBs in hard/quiescent states is $P_{\rm jet}^{\rm XRBs} \sim [L_{\rm jet}^{\rm XRBs}]^{0.85\pm0.06}$, for LLAGN is $P_{\rm jet}^{\rm LLAGNs} \sim [L_{\rm jet}^{\rm LLAGNs}]^{0.71\pm0.11}$. Our results are consistent with theirs, that is, the slopes of jet power and intrinsic $\gamma$-ray luminosity of different types of black hole systems are similar. Our results indicate that all relativistic jets systems may have similar acceleration and emission mechanisms.

Intrinsic $\gamma$-ray luminosity is an indicator of jet power. We study the relationship between intrinsic $\gamma$-ray luminosity and accretion disc luminosity. There is a significant correlation between them. The slope of $\log L^{\rm int} - \log L_{\rm disc}$ of different subsamples is the same within the error range. This result may imply that the formation mechanism of their jets is similar.

### 4.2 Jet power

From our results, we find a significant correlation between jet power and disc luminosity for the whole sample, which supports that jet power has a close link with accretion. If relativistic jets are powered by a Poynting flux, Ghisellini (2006) indicated that the jet power of the BZ can be expressed as

$$L_{\rm BZ} \sim 6 \times 10^{20} \left(\frac{a}{m}\right)^2 \left(\frac{M_{\rm BH}}{\rm M_\odot}\right)^2 B^2 \ {\rm erg\ s^{-1}} \tag{25}$$

where $\alpha/m$ is the angular momentum of a black hole ($\sim 1$ for maximally rotating black holes), and the magnetic field $B$ is in Gauss. Assume that the magnetic energy density near the black hole $U_B \equiv B^2/(8\pi)$ is a fraction of the available gravitational energy $\varepsilon_B$:

$$U_{\rm B} = \varepsilon_{\rm B} \frac{GM_{\rm BH}\rho}{R} = \varepsilon_{\rm B} \frac{R_S}{R} \frac{\rho c^2}{2} \tag{26}$$

where $R_{\rm S}$ is the Schwarzschild radius, $R$ is the stellar radius. The density $\rho$ is linked to the accretion rate $\dot{M}$ through

$$\dot{M} = 2\pi R H \rho \beta_R c \tag{27}$$

where $\beta_R c$ is the radial infalling velocity. There is a relation between mass accretion rate $\dot{M}$ and the disc luminosity

$$L_{\rm disc} = \eta \dot{M} c^2 \tag{28}$$

The BZ jet power can then be written as

$$L_{\rm BZ,jet} \sim \left(\frac{\alpha}{m}\right)^2 \frac{R_{\rm S}^3}{HR^2} \frac{\varepsilon_{\rm B}}{\eta} \frac{L_{\rm disc}}{\beta_{\rm R}} \tag{29}$$

where $H$ is the disc thickness; $\eta$ is the accretion efficiency. The maximum BZ jet power is obtained setting $R \sim H \sim R_S$, $a/m \sim 1$, $\varepsilon_{\rm B} \sim 1$, and $\beta_{\rm R} \sim 1$. In this case

$$L_{\rm BZ,max} \sim \frac{L_{\rm disc}}{\eta} \tag{30}$$

This is in qualitative agreement with what can be estimated in blazars and microquasars, and also in GRBs (Ghisellini 2006). From equation (30), we have

$$\log L_{\rm BZ,max} = \log L_{\rm disc} + \log(1/\eta) + const. \tag{31}$$

According to equation (31), we find that the theoretically predicted coefficient of $\log L_{\rm BZ, max}$–$\log L_{\rm disc}$ relation is 1. Using linear regression, we derive $\log P_{\rm jet} \sim (1.00 \pm 0.02) \log L_{\rm disc}$ for the whole sample, which is consistent with the theoretically predicted coefficient of $L_{\rm BZ, max} - L_{\rm disc}$. Our above results may suggest that the jet power of jetted AGN is powered by the BZ mechanism. Chai et al. (2012) found that the BZ mechanism may dominate over the BP mechanism for the jet acceleration for radio-loud AGNs. Xiong & Zhang (2014) found that there is still a significant correlation between bulk Lorentz factor and black hole mass for Fermi FSRQs, which suggests that jets of Fermi FSRQs are powered by the BZ mechanism. Zhang et al. (2015) suggested that jets of the GeV–FSRQs are launched by the BZ process via extracting the rotational energy of the black hole. Liodakis et al. (2017) studied the relation between radio luminosity and black hole mass from stellar mass to supermassive black hole mass. They pointed towards the Blandford–Znajek mechanism as the dominant mechanism for jet production in black hole-powered jets. Xiao et al. (2022) suggested that jets of BL Lacs are powered by extracting black hole rotation energy, namely the BZ mechanism.

## 5 CONCLUSIONS

In this work, we use a large sample of jetted AGNs to study the relation between jet power, intrinsic $\gamma$-ray luminosity, and accretion. Our main results are the following:

(i) There is a significant correlation between jet power and intrinsic $\gamma$-ray luminosity for the whole sample. The radio galaxies and

$\gamma$ NLS1s follow the $P_{jet}$–$L^{int}$ relation was derived for Fermi blazars. The slope indices we derived are $0.85 \pm 0.01$ for the whole sample, $0.70 \pm 0.02$ for the FSRQs, $0.83 \pm 0.03$ for the BL Lacs, $0.68 \pm 0.11$ for the $\gamma$ NLS1s, and $0.93 \pm 0.09$ for the radio galaxies, respectively.

(ii) There is a significant correlation between intrinsic $\gamma$-ray luminosity and accretion disc luminosity for the whole sample. The $\gamma$ NLS1s and radio galaxies almost follow the $L^{int}$–$L_{disc}$ correlation was derived for Fermi blazars. The slope indices are $1.05 \pm 0.02$ for the whole sample, $0.94 \pm 0.05$ for the FSRQs, $1.14 \pm 0.05$ for the BL Lacs, and $0.92 \pm 0.18$ for the $\gamma$ NLS1s, respectively. Our results support that jet power has a close link with accretion.

(iii) The jet power of almost all sources is slightly larger than the disc luminosity. The jet power depends on both the Eddington ratio and black hole mass. The Eddington ratio and black hole mass have different contribution rates to jet power for jetted AGNs.

(iv) We obtain $\log P_{jet} \sim (1.00 \pm 0.02) \log L_{disc}$ for the whole sample, which is consistent with the theoretically predicted coefficient of $\log L_{BZ, max} - \log L_{disc}$. Our results may suggest that the jet power of jetted AGNs is powered through the BZ mechanism.

(v) The FSRQs and BL Lacs can be divided by using $\log L_{BLR}/L_{Edd} = -0.12 \log L^{int}/L_{Edd} - 2.99$ relation. The $\gamma$ NLS1s fall in the region of FSRQs, while radio galaxies fall in the region of BL Lacs.

## ACKNOWLEDGEMENTS

We are very grateful to the referee and Editor for the very helpful comments and suggestions that improved the presentation of the manuscript substantially. YC is gratefulfor financial support from the National Natural Science Foundation of China (No. 12203028). This work was support from the research project of Qujing Normal University (2105098001/094). This work is supported by the youth project of Yunnan Provincial Science and Technology Department (202101AU070146, 2103010006). YC is grateful for funding for the training Program for talents in Xingdian, Yunnan Province. QSGU is supported by the National Natural Science Foundation of China (No. 11733002, 12121003, 12192220, and 12192222). We also acknowledge the science research grants from the China Manned Space Project with NO. CMS-CSST-2021-A05. This work is supported by the National Natural Science Foundation of China (11733001 and U2031201).

## DATA AVAILABILITY

All the data used here are available upon reasonable request. All data are in Table 1.

This paper has been typeset from a TEX/LATEX file prepared by the author.